# A novel Doppler backscattering (DBS) system to simultaneously monitor radio frequency plasma fluctuations and low frequency turbulence

S. Chowdhury[1, a)], N. A. Crocker[1], W. A. Peebles[1], T. L. Rhodes[1], L. Zeng[1], B. Van Compernolle[2], M. Brookman[2], R. I. Pinsker[2] and C. Lau[3]

[1]*Physics and Astronomy Department, University of California Los Angeles, California 90098, USA*
[2]*General Atomics, P.O. Box 85608, San Diego, California 92186-5608, USA*
[3]*Oak Ridge National Laboratory, Oak Ridge, TN, USA*

a) Author to whom correspondence should be addressed: schowdhury@physics.ucla.edu.

A novel quadrature Doppler Backscattering (DBS) system has been developed and optimized for the E-band (60-90GHz) frequency range using either O-mode or X-mode polarization in DIII-D plasmas. In general, DBS measures the amplitude of density fluctuations and their velocity in the lab frame. The system can simultaneously monitor both low-frequency turbulence (f < 10MHz) and radiofrequency plasma density fluctuations over a selectable frequency range (20-500 MHz). Detection of high-frequency fluctuations has been demonstrated for low harmonics of the ion cyclotron frequency (e.g., $2f_{ci}$~23MHz) and externally driven high-frequency helicon waves (f = 476MHz) using an adjustable frequency down conversion system. Importantly, this extends the application of DBS to a high-frequency spectral domain while maintaining important turbulence and flow measurement capabilities. This unique system has low phase noise, good temporal resolution (sub-millisecond) and excellent wavenumber coverage ($k_\theta$ ~ 1-20cm$^{-1}$ and $k_r$ ~ 20-30cm$^{-1}$). As a demonstration, localized internal DIII-D plasma measurements are presented from turbulence (f ≤ 5MHz), Alfvenic waves (f~6.5MHz), ion cyclotron waves (f ≥ 20MHz) as well as fluctuations around 476MHz driven by an external high-power 476 MHz helicon wave antenna. In the future, helicon measurements will be used to validate GENRAY and AORSA modeling tools for prediction of helicon wave propagation, absorption and current drive location for the newly installed helicon current drive system on DIII-D.

## I. Introduction

Doppler backscattering (DBS)[1-4] is a non-invasive active diagnostic technique that allows the study of electron density fluctuations via backscattering of a millimeter-wave beam launched obliquely towards a cut-off layer. This technique has been used over the past three decades to study the transport role of intermediate scale turbulence as well as determine flow velocity of the turbulence (the $E \times B$ velocity $\tilde{v}_{E \times B}$ can often be extracted from this velocity). The probed wavenumber is controlled by steering the launched mm-wave beam using mirrors and quasi-optical components. When adjusted for normal incidence to the plasma cutoff the technique reduces to well-known reflectometry. Radiation backscattered from perturbations of the index of refraction (typically near cutoff) carries information about $\tilde{n}_e$ (also $\tilde{b}_\parallel$ for X-mode polarized millimeter waves) at the scattering location. The backscattering process is constrained by the Bragg condition $k_{\tilde{x}} = -2k_i$ (where $\tilde{x}$ refers to the probed fluctuation and $i$ to the incident wave at the scattering location), so the DBS technique probes fluctuations with a particular wavenumber, $k_{\tilde{x}}$ which depends in the wave number and incident angle of the incident beam. From the Doppler frequency shift in the backscattered turbulence signal, one can determine the turbulence flow velocity, which can often be converted into a local $E \times B$ velocity. Detection of radio frequency fluctuations can occur in two different ways. The DBS beam can directly backscatter from the radio frequency (RF) wave itself (which can have an associated $\tilde{n}_e$ also $\tilde{b}_\parallel$ at the RF wave frequency) when the RF wave satisfies the Bragg condition. Alternatively, when the RF wave has long wavelength, the probe beam can scatter from low frequency turbulence where the turbulence (or the mm-wave probe beam itself) is modulated by the RF wave.

DBS provides very good measurement resolution (sub-millisecond temporal and sub-cm range spatial) for both the edge and core plasma. This diagnostic has recently been established as a reliable and powerful tool for burning plasma research[5,6]. Historically, DBS has been used to study a wide range of plasma phenomena such as the low to high confinement mode (L-H) transition[7], geodesic acoustic modes (GAMs)[8], zonal flows[9], ion cyclotron emission (ICE)[10], lower hybrid radio frequency waves from an external antenna[11] etc. The current paper reports on the design and initial results from a novel quadrature DBS system performance over a wide signal frequency range f~0–500 Mhz. Section IV discusses and summarizes the important measurement results obtained using a system prototype during recent helicon current drive experiments on DIII-D.

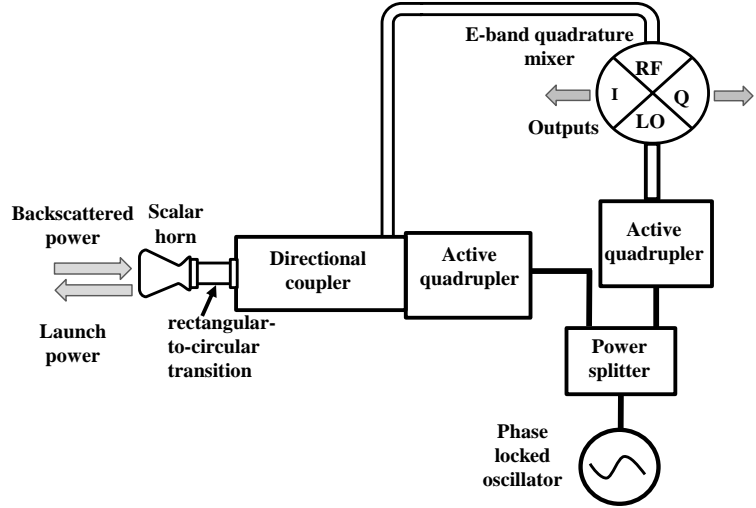

FIG. 1. A circuit for millimeter wave probe beam production and DBS measurement.

system developed for application on the DIII-D[12-15] tokamak in the E-band frequency range (60-90GHz) to simultaneously monitor low frequency (LF, f≤10MHz) turbulence and high frequency (HF, f≥20MHz) fluctuations that can extend to the radio frequency (RF, f~500MHz) range. The main advantages of this DBS system are the sensitivity, very low phase noise and radio frequency measurement flexibility over a wide range of frequencies while providing the capability of simultaneously monitoring low frequency turbulence and plasma flow. This allows investigation of any features of the high frequency measurement resulting from the simultaneous presence of low frequency turbulence, or vice versa. The primary aim of this novel diagnostic is to demonstrate its capability to measure helicon wave plasma fluctuations and test theory and simulations. This will help to advance the objectives of the helicon current drive program[16-18] at DIII-D, which is to establish the physics basis for off-axis current drive with helicon waves (aka fast waves in the whistler regime) in the high-beta reactor-relevant plasmas. A DBS system will provide measurements of helicon wave-field amplitude and spatial structure in the core of plasmas with high-power helicon injection in order to validate codes that predict mode propagation and absorption of the helicon waves. It can also potentially aid in diagnosing unwanted coupling of power from the helicon antenna into the slow wave mode in the scrape off layer (SOL)[19].

Section II describes the details of the system design. Laboratory testing and preliminary DIII-D plasma results are described in Sec. III. The plasma results demonstrate the

## II. SYSTEM DESIGN

The heart of the new DBS system is the millimeter wave circuit, shown in Fig. 1. The millimeter wave circuit is driven by a fixed frequency phase-locked dielectric resonant oscillator (PLDRO). PLDROs have very low phase-noise (< ~120dBc/Hz at 1MHz) and high long-term stability, as well as insensitivity to the background neutron radiation expected in the vicinity of DIII-D during hot plasma experiments. By design, the system can employ sources in the range 15 – 22.5GHz. For the plasma test results reported in Section III, two different PLDROs, at 15.75 and 18GHz, are utilized at different times, dependent on the plasma profile parameters. As shown in Fig. 1, the output power from the chosen source is equally divided using a Wilkinson power splitter (Krytar, 6020265). The outputs pump two separate E-band active quadruplers (Eravant, SFA-603903420-12KF-E1). One provides the launch power, which is directed to a target (a reflector, for lab testing, or the DIII-D plasma). The other multiplier delivers the required local-oscillator (LO) power (16dBm) for the millimeter-wave quadrature mixer (Eravant, SFQ-60390315-1212SF-N1-M). A fixed waveguide attenuator (not shown in figure) regulates the LO power for optimum operation of the quadrature mixer. After the quadruplers, WR-12 waveguide is used for wave transmission. The launch power is directed to the target via a 3dB directional coupler (Eravant, SWD-0340H-12-SB), rectangular-to-circular transition, scalar horn and a quasi-optical steering optics (not shown in the figure). The backscattered received power is mixed with the LO in the E-band quadrature mixer.

The de-modulated outputs provide the in-phase (I) and quadrature (Q) information that allows determination of the received electric field amplitude and phase variations over time.

The quadrature mixer outputs are processed by either of two different receiver circuits (Figs. 2 and 3) for the shunted to common to minimize signal contamination by local pickup of low frequency electromagnetic interference.

For initial DIII-D plasma tests, the design of the receiver circuit (Fig. 2) is optimized for relatively low frequency RF waves (e.g., energetic-ion driven ICE, typically observed at $f \geq 20$MHz in DIII-D[10]). As shown in

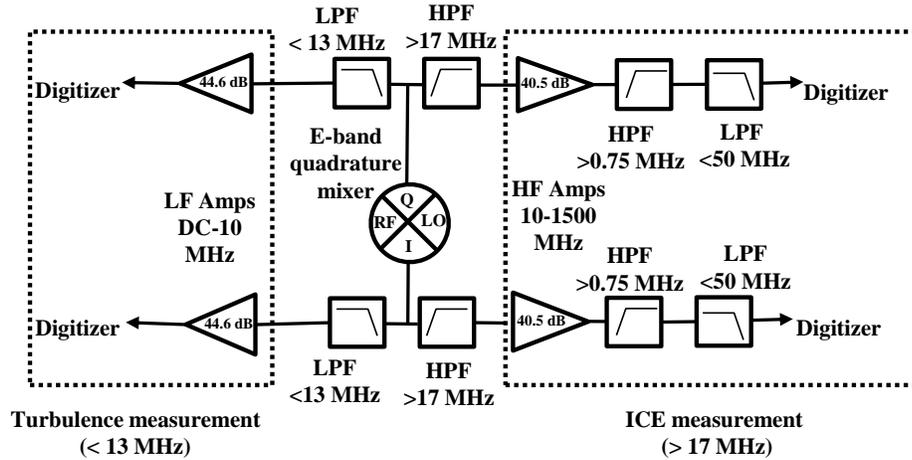

FIG. 2. DBS receiving electronics for low frequency (LF) and high frequency (HF) fluctuations.

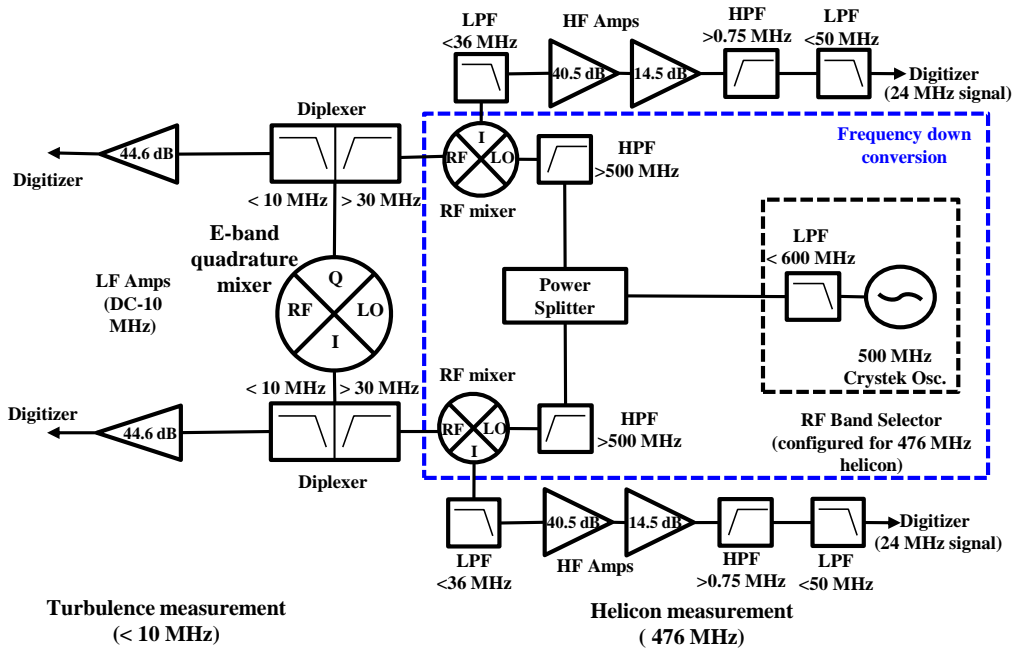

FIG. 3. Receiving circuit with frequency down conversion setup for high frequency fluctuation measurement in helicon frequency range (f = 476 MHz). A 500 MHz crystal and RF mixer (positioned close to the mm-wave mixer) down-convert the detected high frequency fluctuation to a lower range (476 MHz to 24 MHz).

results presented in Section III. For both receiver circuits, all active millimeter wave components, amplifiers and sources are powered with regulated supplies for stable operation and power input terminals are capacitively

Fig. 2, the quadrature mixer output signals (both I and Q) are each separated into low frequency (LF) and high frequency (HF) bands via low pass ($\leq$13MHz) and high pass ($\geq$17MHz) filters. (Min-Circuits SLP-15+ and SHP-

20+, respectively, are used. Both have 3dB cutoffs at ~ 15MHz, individually. When joined together via a tee, as in Fig. 2, there is some overlap at ~ 15MHz.) This division into low and high frequency components allows the LF and HF signals to be amplified separately, which is useful because the HF fluctuations tend to have a much smaller amplitude than the LF turbulence. The amplifier gain for the HF signal is ~ 40.5dB (Mini-Circuits ZKL-1R5+). A low noise 44.6dB gain amplifier (custom-made) is used for the LF signals. Various attenuators (not shown in Fig. 2) are utilized to prevent saturation of the LF signal amplifers and digitizer, leading to a significantly lower net gain for the LF signals than the HF signals. The amplified LF and HF components of the I and Q signals are also both filtered with a 50MHz low-pass anti-alias filter and recorded by a multichannel digitizer at 100MS/sec. (The digitizer model is the Alazartech ATS9416, featuring 14-bit resolution). The HF component is also further high-pass ($f \geq 750kHz$) filtered at the digitizer input to eliminate cable pickup of low frequency electromagnetic interference (Mini-Circuits ZFHP-0R75-S+).

For measurement of the 476MHz helicon wave, the receiver circuit is modified as shown in Fig. 3 to down-convert the high frequency components of the I and Q signals using a 500MHz local oscillator (LO) before amplification, translating the detected helicon signal to 24MHz. Given the high power of the helicon wave injected into the plasma ($\lesssim 1$ MW), down-conversion is essential to reduce unwanted electromagnetic pick-up of stray 476 MHz radiation outside the tokamak. The down-converting RF mixers are positioned physically close to the millimeter-wave quadrature mixer (within a few inches, limited by the physical dimensions of the diplexers) to minimize pickup. Other advantages of down-conversion are that it allows for low-cost digitization of the signal and eliminates significance cable loss that would occur at the helicon frequency. In addition, the use of different LO down-conversion frequencies allows measurement of RF waves over different frequency ranges. The processing of the low frequency components is unchanged from the non-down-converting receiver circuit in Fig. 2.

As shown in Fig. 3, diplexers (Mini-Circuits model ZDPLX-2150) are used to separate the signals from the quadrature mixer into low and high frequency components. The down-conversion is accomplished using radio frequency mixers (Mini-Circuits ZFM-1W-S). The LO is supplied by a 500MHz crystal oscillator (Crystek RFPRO33 500.000M). A 600MHz low pass filter (Min-Circuits SLP-600) rejects any higher harmonic content in the crystal oscillator signal. Another difference from the receiver circuit of Fig. 2 is the second stage of amplification with ~ 14.5dB gain (Mini-Circuits ZX60-P105LN+) for the high frequency channel giving a total amplifier gain of ~55dB to the down-converted signal.

Several measures are taken to prevent unwanted 476MHz pickup from getting into the down-converting RF mixer. High pass filters (KR-electronics Inc. KR-3463-SMA) with a sharp cut-off below 500MHz are placed directly at the LO ports of the RF mixers (~30dB rejection at 476MHz). The crystal oscillator is powered by a battery coupled via a 5MHz low pass filter (LPF) (Mini-Circuits SLP-5). Finally, bi-directional 36MHz low pass filters (Mini-Circuits SLP-36+) are directly connected to the IF outputs of the RF mixers to prevent helicon pickup from entering via the IF output ports.

The system is tested in the laboratory, and then later installed on the DIII-D tokamak and tested during plasma experiments, using both receiving circuits (Figs. 2 & 3). The system design and component arrangement are optimized for compactness and portability, to facilitate easy installation at DIII-D and easy relocation to different ports offering access to the plasma. The millimeter wave circuit and receiver circuit, along with power supply regulators are installed on a portable 12″ × 18″ optical breadboard. Results of the plasma tests are reported in Section III.

As noted above, the down-converting receiver of Fig. 3 can be easily modified to target RF plasma waves at different frequencies. This capability is exploited for some of the test results reported in Section III targeting lower harmonics of ICE. To target these plasma waves, the circuit of Fig. 3 is modified by replacing the 500MHz crystal oscillator with a 50 MHz crystal oscillator (Crystek CHPRO 033 50.000), and the $f < 600MHz$ low pass filter on the output of the crystal oscillator is replaced with a $50 < f < 71MHz$ bandpass filter (Min-Circuits SBP-60) for harmonic rejection. For convenience, this modified version of the circuit in Fig. 3 will be referred to throughout this paper as the "modified down-converting circuit". Also, the assorted filters described above for 476MHz pickup rejection are removed. This includes, (in particular) $f < 36MHz$ low pass filters on the RF mixer (down-converting) output ports, since they would have the side effect of restricting the range of sensitivity of the ICE measurement. No pickup rejection filters for ICE are added in their place. Since ICE is a low power plasma wave driven by an instability, there is little danger of contamination of the measurement by electromagnetic interference with the measurement circuit due to stray radiation outside the tokamak.

For the laboratory tests, a collimated millimeter-wave beam is launched using the scalar horn, as shown in Fig. 1, and an aspheric lens (f = 10"), towards a flat reflector approximately ~ 1.5m from the lens, where the beam is retroreflected. The mirror is translated a distance of a few inches along the beam to introduce a path length variation in the retroreflected radiation and a change in phase relative to the local oscillator. The path length variation manifests as variation in the I and Q signals from the E-band quadrature mixer, which are related to the amplitude (A) and phase ($\phi$) of the retro-reflected radiation by $I = A\cos(\phi)$ and $Q = A\sin(\phi)$. The amount of retroreflected power in this arrangement is much larger than the mixer can accept, so an absorber is attached to the surface of the reflector to reduce the retroreflected power. A clean circular phasor in the plot of $I$ vs $Q$ from the low frequency signal amplifiers is observed as the reflector is translated, as expected for normal operation. In addition, using the PLDROs phase noise is found to make a negligible contribution to the overall system noise which is set primarily by amplifier and mixer noise.

The CPS interface couples radiation from a scalar horn to the plasma by imaging radiation from the horn via a series of quasi-optical elements into an overmoded 3.5" diameter (ID) corrugated waveguide which penetrates the vacuum vessel. Radiation from the waveguide is focused by a metallic dichroic lens[22] onto a flat steering mirror. A two-axis motorized system is attached to the steering mirror to control beam launch angle both poloidally (-30° to 20°) and toroidally (-10° to 5°) in order to allow coverage of a wide range of possible wave numbers at cutoff. The poloidal steering enables the DBS beam to cover a wave number

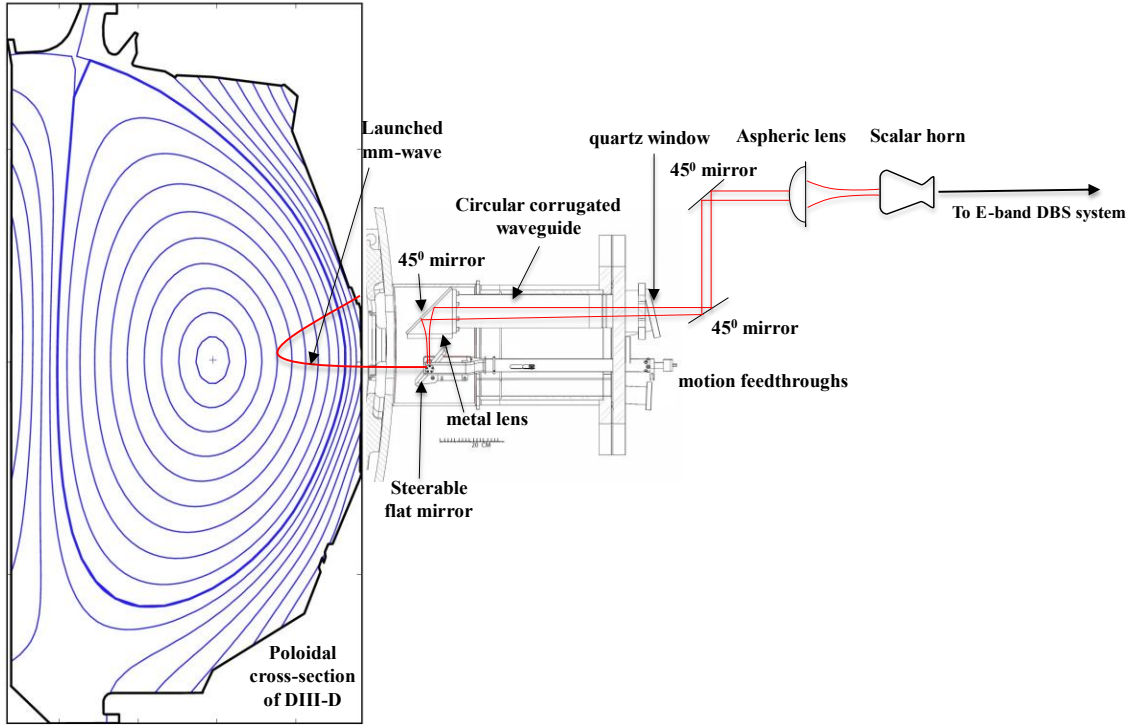

FIG. 4. DBS installed at DIII-D near mid plane at low field side borrowing existing V-band cross polarization optics[20-22]. A waveguide transition from E → V is used for 63 or 72GHz microwave launch. A collimating aspheric lens (made of HDPE) and a combination of two plane mirrors are used to collimate beam to in-vacuum quasi-optics. The vacuum quasi-optics mainly contains 3.5" long corrugated waveguide, 45⁰ plane mirror, metallic lens and a steerable flat mirror for beam steering. A two-axes motorized system is attached to this mirror in vacuum to steer (remotely) the mirror at different poloidal as well as in toroidal angle to target different cut-off locations. Back-scattered plasma signal is then collected via same scalar horn antenna finally to the receiving electronics. A polarizer has been used (not showed in fig) to select either X/O-mode polarization for the launched DBS beam. Left hand side shows a set of blue contour lines indicating magnetic flux surfaces bounded by a material plasma facing surface. The plasma is largely confined to the region of closed flux surfaces, extending slightly into the region beyond the last closed flux surface referred to as the scrape-off layer (SOL). The red line shows a typical millimeter wave beam trajectory for the DBS system.

For the DIII-D test results described in Section III, the system is installed near the midplane on the low magnetic field side of the DIII-D tokamak. Taking advantage of its compact, portable configuration, the prototype is coupled to the plasma for these tests by integrating the system into an existing plasma interface developed for a V-band cross-polarization scattering (CPS) system described in Ref.[20-22]. The transition and scalar horn shown in Fig. 1 are replaced by an E-band-to-V-band transition in order to couple into the V-band transmission system of the interface. The two PLDRO sources used for results reported in Section III produce launch frequencies of 63 or 72GHz, which are within the V-band transmission system.

range of $k_\theta \sim$ 1-20cm$^{-1}$ at cutoff, while the toroidal steering allows minimization of wave number mismatch for probing plasma waves of interest (e.g., minimization of $k_\parallel$ for probing turbulence[23]). This matching requirement becomes more important as the probed wavenumber increases. Notably, the beam can also cover $k_r \sim$ 20-30cm$^{-1}$ in the edge, (as in ref. [11]) since backscattering can occur anywhere along the DBS beam if the Bragg scattering conditions are met: $\omega_i + \omega_{\tilde{x}} = \omega_s$ and $\boldsymbol{k_i} + \boldsymbol{k_{\tilde{x}}} = \boldsymbol{k_s}$ where $\omega$ and $\mathbf{k}$ correspond to frequencies and wave number vectors, the subscripts i and s correspond to the incident and scattered millimeter waves and $\tilde{x}$ to the plasma fluctuation causing the scattering. Normally, $\omega_i \approx \omega_s \gg \omega_{\tilde{x}}$, $k_s \approx -k_i \approx \frac{1}{2}k_{\tilde{x}}$ are satisfied by a plasma wave (e.g., helicon or slow waves at 476MHz launched by the helicon antenna[24]). The steering angles are chosen and set prior to creation of the discharge in order to probe the desired wave number. The choice is made using millimeter wave ray tracing (GENRAY[25]) for a reference discharge that is taken to be a model for the upcoming discharge.

The CPS interface includes a capability to switch between X- and O-mode polarization, allowing millimeter waves to be launched into the plasma with either polarization and returning millimeter waves of the same polarization to be received. This capability enables variation in measurement location by choice of operating polarization. The switching is accomplished by routing launch power to either of two horn antennae that are roughly oriented for X- or O-mode polarization and the use of a wire-grid polarizer to fine-tune the polarization. The exact choice of polarization angle is made prior to creation of the discharge with the goal of minimizing coupling to the unwanted mode, considering the expected pitch of the edge magnetic field, which is typically determined from the equilibrium reconstruction of the plasma discharge.

## III. SYSTEM PERFORMANCE AND INITIAL PLASMA MEASUREMENTS

The DBS system is installed on the DIII-D tokamak for testing via plasma measurements. These initial tests are performed using plasmas with current and toroidal magnetic field of $I_P = 1 - 2$MA and $B_T = 1.5 - 2$T, respectively with injected neutral beam power as high as ~10MW. Measurements are obtained for both high-confinement (H-mode) plasmas, which tend to have broad density profiles, and low-confinement (L-mode) plasmas, which tend to have more centrally peaked density profiles. Measurements are obtained using both sources (i.e., for launched frequencies of 63 or 72GHz) and a broad range of poloidal (-18° to 0°) and toroidal (-7° to 0°) angles allowing wavenumbers of $k_\theta$ ~ 1-10cm$^{-1}$ to be probed. The following results are selected to demonstrate the system performance at low frequency (f ≤ 10MHz), intermediate frequency (f ≥ 20MHz), and high frequency (i.e., the helicon system frequency range f ~ 476MHz).

Initial testing using the non-down-converting receiver circuit of Fig. 2 demonstrates the capability of the system to measure both turbulence and Alfvén eigenmodes. Figure 5 shows O-mode measurements with a launch frequency of 63GHz in a neutral beam heated ($P_{NB}$ ~ 8MW) H-mode plasma with central magnetic field $B_T = 2$T and plasma current $I_P = 1.3$MA (shot # 186656). The central electron density is $n_{e0}$ ~ 6.25 x 10$^{13}$cm$^{-3}$ for the time period shown (t = 2500 – 2700ms). The poloidal steering angle for these measurements is -10.2°, resulting in a wave number of $k_\theta =$ 2.3cm$^{-1}$ probed at cutoff at $\rho \approx 0.7$. (The quantity $\rho$ is a radial label for the closed magnetic flux surfaces in the plasma - cf. Figure 4 - ranging from 0 to 1 between the magnetic axis and the last closed flux surface. For each surface, $\rho$ corresponds to the square root of normalized toroidal flux enclosed by the flux surface.) The toroidal steering angle for these measurements is 0°. Fig. 5 shows spectrograms of quadrature electric field fluctuations from the low-frequency output of the system with log$_{10}$ scale in intensity.

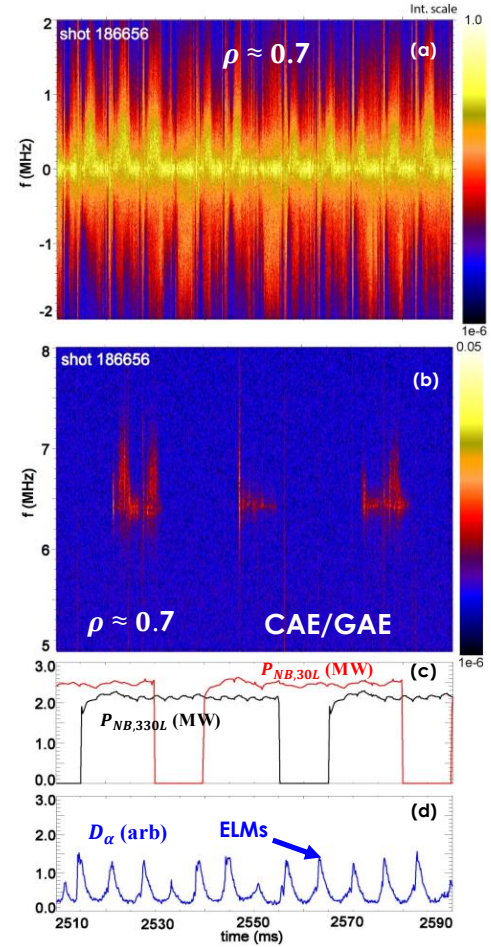

FIG. 5. Two different frequency bands of quadrature spectrum for low frequency signal are shown in (a) and (b) for receiver circuit as shown in Fig. 2 using 15.75 GHz DRO as a low frequency source to launch 63GHz mm-wave in O-mode polarization. (a) Low frequency turbulence in frequency band $f$ = -2MHz to +2MHz and (b) high frequency CAE/GAE mode at 6.5 MHz in band $f$ = 5 to 8MHz. (c) Injected beam power for two different co-injecting beams ($P_{NB,30L}$ and $P_{NB,30L}$) and (d) $D_\alpha$ optical emission. Jumps in $D_\alpha$ correspond to ELMs. Turbulence spectrum in panel (a) shows modulation correlated with ELMs, while CAE/GAE mode in panel (b) appears (after a ~ 6ms delay) during periods when both beams are simultaneously injecting at full power.

The quadrature electric field is proportional to $\tilde{n}$ for a scattering measurement. Fig. 5a shows turbulence in the range $|f| \leq 2$MHz and Fig. 5b shows high frequency Alfvén eigenmodes, of either compressional[26] or global[27] polarization, at ~ 6.5MHz ~ 0.53$f_{ci}$ where, $f_{ci}$ is the ion cyclotron frequency at the location of the millimeter wave cutoff. The turbulence in Fig. 5a shows a clear modulation of Doppler shift and intensity. This is correlated with the occurrence of edge localized modes (ELM), which are indicated by bursts of $D_\alpha$ emission (Fig. 5d). The observed modulation of the turbulent spectrum is at least partly attributable to a change in measurement location since the ELMs cause significant transient changes to the edge density profile. The short periods of Alfven eigenmode activity in Fig. 5b occur whenever both the 30L and 330L beams are simultaneously on, indicating the AEs are excited by fast ion population created by the combination of these beams, which both have the same tangency radius and input beam voltage (~80kV) and thus create similar fast-ion populations. This suggests that the combined beams cause the fast-ion density to rise above a stability threshold whereas one beam by itself is insufficient to drive the CAE/GAE[27]. The modes disappear as soon as either of the beams (30L/330L) is turned off, whereas during periods where both beams are on, a ~6ms delay is observed for mode appearance after the moment both beams become simultaneously on. The reason behind this delay requires further analysis and measurement, which is beyond the scope of this paper. In addition, there is a difference of ~100kHz between the magnitudes of the frequencies of the peaks seen in the negative (not shown) and positive (Fig. 5b) frequency sides of the spectrum for the observed mode. This difference is approximately twice the Doppler shift $f_{Doppler}$ ~ 60kHz (t = 2550ms) in the turbulence spectrum. This indicates that the peaks in the high frequency spectrum are caused by modulation of the turbulence-scattered radiation (or of the turbulence itself) by long wavelength high frequency modes, as discussed in Ref.[10]. In principle, the peaks in the high frequency spectrum should occur at $f_{Doppler} \pm f_{CAE/GAE}$, so the observed difference frequency should be $2f_{Doppler}$ ~ 120kHz. In practice, the broadening of both the high and low frequency spectral peaks make the determination of the peak frequencies imprecise. The turbulence-modulation interpretation of the high frequency spectrum is further supported by moments of transient broadening and shifting of the high frequency peak (e.g., at t ≈ 2522 ms and t ≈ 2528 ms) that correlate with the ELM-associated broadening and Doppler shift change of the turbulent spectrum.

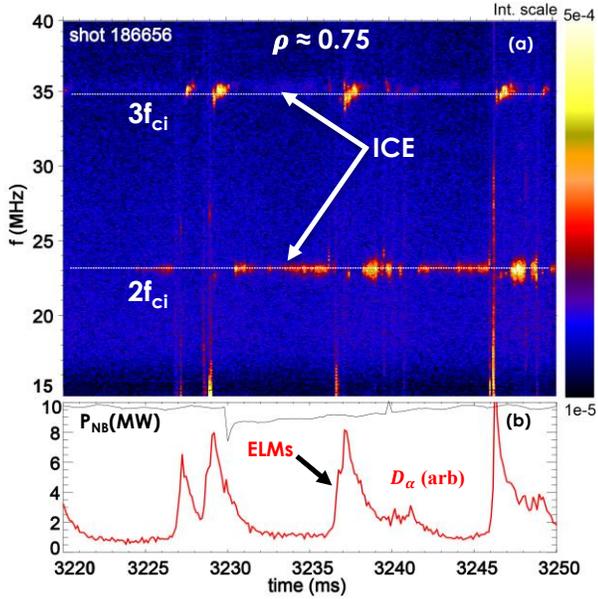

FIG. 6. (a) Quadrature spectrum of high frequency signal using same receiver circuit as shown in Fig. 2 using 15.75GHz DRO as a low frequency source to launch 63GHz mm-wave in X-mode polarization. (b) Total injected beam power ($P_{NB}$) and $D_\alpha$ optical emission. Jumps in $D_\alpha$ correspond to ELMs. ELM-correlated broadband bursts and ELM-related modulation of ICE can be seen in spectrum in panel (a).

Later in the same discharge (shot#186656, t = 3200 – 3250ms), beam-driven low-harmonic ICE is observed in the spectrogram of the high-frequency output of the DBS system (Fig. 6). Neutral beam heating power is increased to $P_{NB}$ ~ 10MW at t = 3000ms. Also, the central density is increased to $n_{e0}$ ~ 6.5 x $10^{13}cm^{-3}$, which moves the cutoff out to $\rho \approx 0.75$, where the system probes $k_\theta$ ~ 2.5cm$^{-1}$. Fig. 6(a) shows beam-driven ICE at harmonics of the edge ion cyclotron frequency ($f$ = 23 MHz ~ $2f_{ci}$, and $f$ = 34.5MHz ~ $3f_{ci}$) appearing at t = 3225ms, shortly after the neutral beam injection power increases. Modulation of the ICE is observed to correlate with ELMs (similar to observations reported in ref.[10]). Short bursts of broadband emission covering the entire measured frequency range are also observed at the ELM crashes. (The short period in which $D_\alpha$ emission sharply rises is referred to as a "crash" because just inside the last-closed flux surface of the plasma, electron density and temperature drop abruptly during this period.) Note that there are significant differences in the evolution of intensity in the peaks at the 2nd and 3rd harmonic ICE radiation in Fig 5a, consistent with interpretation of these spectral features as being associated with plasma waves (as opposed to harmonic artifacts produced by systematic nonlinear effects in the electronics).

For later testing in DIII-D, the down-converting receiver circuit of Fig. 3 is used for measurement of higher frequency fluctuations. To begin with, the receiver circuit of Fig. 3, modified for ICE as described in Section II, is tested using ICE as the target wave. Measurements of edge density fluctuation are obtained for a beam heated (~7MW) H-mode discharge using a 63GHz launch frequency but in X-mode polarization. Fig. 7(a) shows a clear signature of 3rd harmonic ICE at $3f_{ci}$~34.5MHz in the density fluctuation spectrograph measured near the plasma edge ($\rho \approx 0.95$). Note that the actual recorded frequency is ~15.5MHz. As discussed in Section II, this potentially corresponds to a plasma wave at either ~34.5MHz or ~65.5MHz, considering the 50MHz frequency shift. The measurement is interpreted to correspond to a plasma wave at ~34.5MHz, and the figure frequency axis is adjusted accordingly, since ICE is expected to be stronger at lower harmonics. A frequency of ~34.5MHz would correspond to the 3rd ICE harmonic at plasma edge, $\rho \sim 1$, whereas ~ 65.5MHz would match the 5th ICE harmonic deep in plasma core, at $\rho \sim 0.4$. This interpretation is also consistent with other DBS measurements of ICE presented here and in ref.[10], which show frequencies matching harmonics in the edge, not the core. At t = 1890ms, the discharge begins to exhibit ELMs, as can be seen from the $D_\alpha$ trace in Fig. 7b, and the 3rd harmonic disappears nearly completely from the spectrum. This is potentially because of fast-ion transport caused by ELMs[10]. These results confirm the down-conversion receiver circuit sensitivity to ICE harmonic and motivates its usefulness in measuring waves at higher frequencies than ICE.

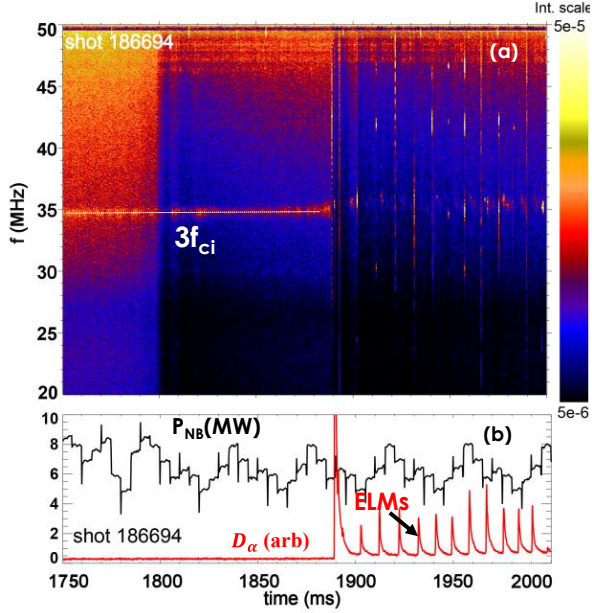

FIG. 7. (a) Quadrature spectrum in density contour ($\log_{10} \tilde{n}$ scale) for high frequency channel shows 3rd harmonic edge ICE ($\rho \sim 0.95$). (b) A neutral beam of $P_{NB} \sim 7$MW is injected and ELM appears at ~1890ms. DBS mm-circuit uses frequency down-conversion (as described in Fig. 3) using 50 MHz crystal to test the ICE sensitivity. ICE harmonic disappears after $t \sim 1890$ms due to ELM effects on fast ion transport.

After completing tests of the ICE-modified down-converting receiver, the receiver is reconfigured for helicon measurement as in Fig. 3 for testing during experiments with high power helicon injection. The DBS with helicon down-converting receiver circuit is tested during high power (~450kW) helicon injection during both low-confinement (L-mode) and high-confinement (H-mode) discharges. A 72GHz millimeter-wave beam is launched in X-mode to probe helicon fluctuations near outer midplane, in the edge plasma.

Fig. 8 shows measurements for an L-mode plasma with $n_e \sim 2 \times 10^{13} cm^{-3}$ where a short pulse of ~450kW helicon power (Fig. 8c) is injected. The DBS is system steered to probe density fluctuations ($\tilde{n}$) with $k_\theta \sim 3.6$cm$^{-1}$ at the cutoff, which is at $\rho = 0.46$ (the toroidal steering angle is 0°). Quadrature spectra for both low (Fig 8a) and high frequency signals (Fig. 8b) are shown in Fig. 8. A Doppler shift of $f_D \sim 600$kHz of the turbulence spectrum can be observed in the LF quadrature spectrum (Fig. 8a).

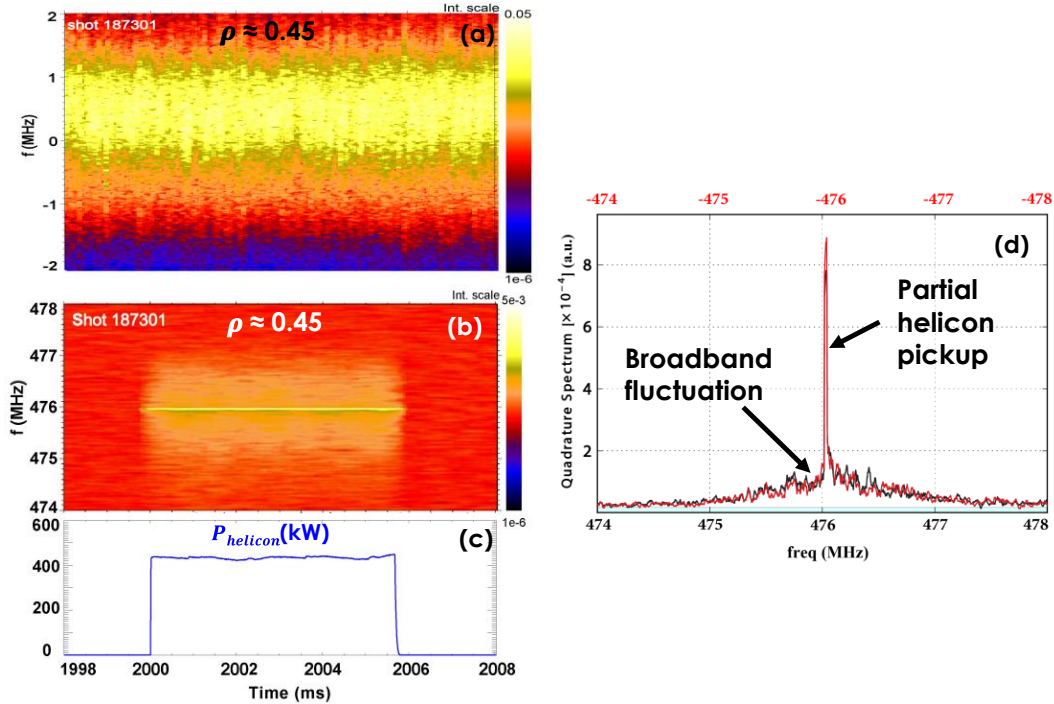

FIG. 8. DBS observes broadband helicon fluctuations during high power (~450kW) helicon injection in L-mode plasma. System uses down conversion setup with 500MHz crystal (as shown in Fig. 3) to down-convert the 476MHz helicon fluctuation to lower frequency (24MHz). Contour of plot quadrature electric field spectrum in $\log_{10}$ - scale shows (a) Doppler shifted ($f_D \sim 600$kHz) turbulence and (b) high-frequency fluctuations around helicon frequency, consisting of a sharp peak at the helicon frequency (476MHz), partially due to pickup, and broadband fluctuations within a band a few MHz wide around 476 MHz. [Int. scales of (a) and (b) differ by an arbitrary factor.] (c) Helicon power vs time. (d) Overlay of positive and negative frequency side of high frequency quadrature spectrum (t=2001ms).

Simultaneous measurements of helicon fluctuations in the HF are shown in Fig. 8(b). A contour plot of the HF fluctuation spectrum shows broadband fluctuations a few MHz wide ($\Delta f \sim 2$MHz) around the helicon injection frequency ($f = 476$MHz), with a relative amplitude ~70dB lower than the simultaneously measured lower frequency turbulence (f ≤ 2 MHz). A sharp peak is also seen in the spectrum at 476MHz, as a sharp yellow line in the spectrum contour plot in Fig. 8b, as well as a sharp peak in time-slice spectra in Fig. 8d. (Note that the frequency scales are adjusted to account for the frequency shift of 500MHz caused by down-conversion. The measured frequency of this peak before adjusting for the down-conversion is actually 24MHz).

The broadband fluctuations are due to measurement of high frequency density fluctuations ($\tilde{n}$) in the plasma, while the sharp peak may be partially due to stray radiation from the helicon antenna picked up by the DBS electronics. Measurements (not shown) are performed during a similar plasma in which the millimeter waves are blocked from entering the plasma. A similar sharp peak at 476MHz is observed, but there are no broadband fluctuations observed around the peak, in contrast with Fig. 8b. The broadening of the spectrum around the helicon frequency is probably due to the interaction of helicon waves with turbulence. One possible form of interaction is that turbulent density fluctuations in the plasma edge near the helicon antenna modulate the amplitude of the helicon wave coupled from the antenna to the plasma. The injected helicon wave is evanescent in the vicinity of the antenna where density is low, and only propagates freely deeper inside the plasma where density is higher. Amplitude modulation would manifest as a symmetric broadening of the spectrum. Full-wave helicon modeling[28] indicates that edge turbulence can significantly modify coupling and propagation in the plasma.

Comparison of the broadband fluctuations on the positive and negative sides of the high frequency quadrature spectrum (shown for, e.g., $t = 2001$ms in Fig. 8d) leads to the preliminary conclusion that the observed broadband fluctuations are the product of direct backscattering of the incident probe beam millimeter waves from plasma waves with frequencies near 476MHz. As discussed in ref.[10], several different physical processes could, in principle, cause backscattered millimeter waves to exhibit the broadband fluctuations seen in Fig. 8b. The first is that the incident millimeter waves could directly backscatter from the turbulence in the presence of the plasma waves. The plasma waves could then introduce a high frequency component into the spectrum of scattered millimeter waves by either modulating the turbulence (via oscillatory ExB motion caused by the plasma wave E-field) or by modulating the index of refraction along the path of the probe beam and backscattered radiation. A second process is that incident millimeter waves could directly backscatter from plasma waves with the observed frequencies if the plasma waves satisfy the Bragg scattering condition for wave number, $\mathbf{k}_{\tilde{x}} = -2\mathbf{k}_i$, where $\mathbf{k}_{\tilde{x}}$ and $\mathbf{k}_i$ are the plasma wave and millimeter wave wavenumbers, respectively. (The plasma wave must have an associated density fluctuation or, if the millimeter wave is X-mode polarized, an asscociated fluctuation in the B field strength.) The comparison of spectra around |f| = 476MHz on the negative and postitive sides of the quadrature spectrum can be used to help distinguish which process is responsible. For the 1$^{st}$ process, scattering from turbulence, the broadband fluctuations should be observed to center around peaks at $f_{Doppler} \pm 476$MHz, where $f_{Doppler}$ is the turbulent Doppler shift frequency at the scattering location. Most of the power scattered from turbulence comes from the probe beam at cutoff, and the simultaneously measured turbulence has a Doppler shift $f_{Doppler} \sim 600$kHz (Fig. 8a), so any broadband fluctuations produced by this process should center around 476.6MHz and -475.4MHz.

In contrast, for the 2$^{nd}$ process, direct backscattering from the plasma waves, the broadband fluctuations on the negative side should center around $f = -476$MHz, since the broadband fluctuations on the positive side of the spectrum center around $f = 476$MHz. From Fig. 8d it can be seen that the broadband fluctuations on the positive and negative side are, in fact, centered around +476MHz and -476MHz, respectively, consistent with direct backscattering of the incident millimeter waves from plasma waves at frequencies close to 476MHz. In principle, backscattering can occur anywhere along the millimeter wave probe beam path. Further modeling is required to determine where along the path the Bragg scattering condition might be satisfied by waves launched by the helicon antenna.

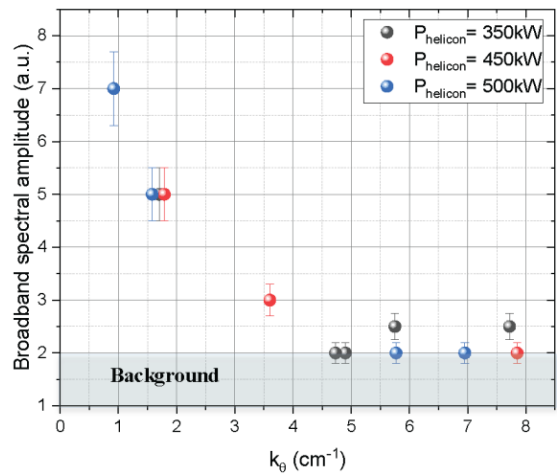

FIG. 9. Broadband fluctuations amplitude around helicon frequency only observed at low $k_\theta$ i.e., at lower scattering angle (within $\pm 5^0$ of normal incidence) for L-mode plasma.

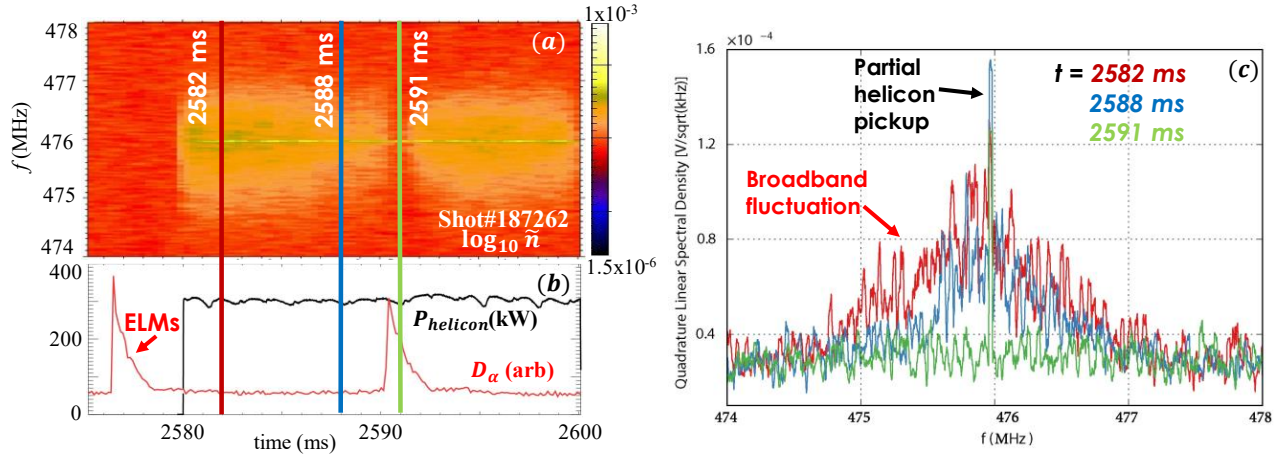

FIG. 10. Helicon broadband fluctuation and its evolution in ELM'y H-mode plasma. 72GHz DBS beam with X-mode polarization is launched at low-k to test helicon antenna power coupling (absorption) by the high-$\beta$ beam heated ($P_{NB} \approx 2.5$MW) plasma. (a) helicon broad-band fluctuation amplitude changes with ELM and completely disappear at ELM event. (b) ~300kW helicon power in pulse mode was injected to H-mode plasma, lower divertor $D_\alpha$ visible light emission shows ELM occurrence at the edge. (c) Partial helicon pickup and helicon broadband fluctuation change during ELM (t = 2582, 2588 and 2591ms) shows real plasma effect on the helicon wave.

A poloidal angle scan with the DBS system over a range of similar L-mode plasmas shows that the broadband fluctuations are observed when the DBS probes fluctuations with $k_\theta < \sim 4$cm$^{-1}$. The cutoff location for this scan is $\rho \sim 0.45$. Fig. 9 shows average spectral amplitude within the broadband range of frequencies (excluding the sharp peak at 476 MHz) vs. probed $k_\theta$. The poloidal angle is varied ±5° around normal incidence (at a toroidal angle of 0°). This dependence of broadband power on $k_\theta$ supports the interpretation that the broadband fluctuations are the result of milllimeter wave scattering from plasma waves, as opposed to being the result of pickup. This also supports the interpretation that the broadband fluctuations are from direct backscattering of the probe beam from helicon plasma waves, suggesting in particular that the scattering takes place near the probe beam cutoff, since the probe beam scatters from waves with $k_\theta \sim 1 - 4$ cm$^{-1}$ at the cutoff. Note that the broadband amplitude shows very little dependence on the amount of helicon injected power, which ranges from 350 to 450kW during this scan. This is consistent with the narrow range of launch powers and the expectation that the amplitude of helicon waves launched from the external antenna should scale as the square root of launched power.

Measurements during a neutral beam heated ($P_{NB}$ ~2.5MW) H-mode discharge show that the broadband fluctuations significantly change during the period around an ELM. Fig. 10a shows quadrature spectra vs time (with 0.5ms temporal smoothing) for the high frequency signal during the time period around an ELM, during injection of 300kW of helicon power (Fig.10b). The broadband fluctuations are seen to decrease during the ~ 5ms period before the crash and then reappear almost immediately after the crash. Figure 10c shows spectra for three time slices around the ELM crash, t = 2582, 2588 and 2591ms, to facilitate visualisation of the spectral evolution. The spectra at t = 2582ms and 2588ms, which are both before the crash, show the decrease in broadband fluctuations over time. The spectrum at t = 2591ms, immediately after the crash, shows that reduction of the broadband fluctuations to the noise level.

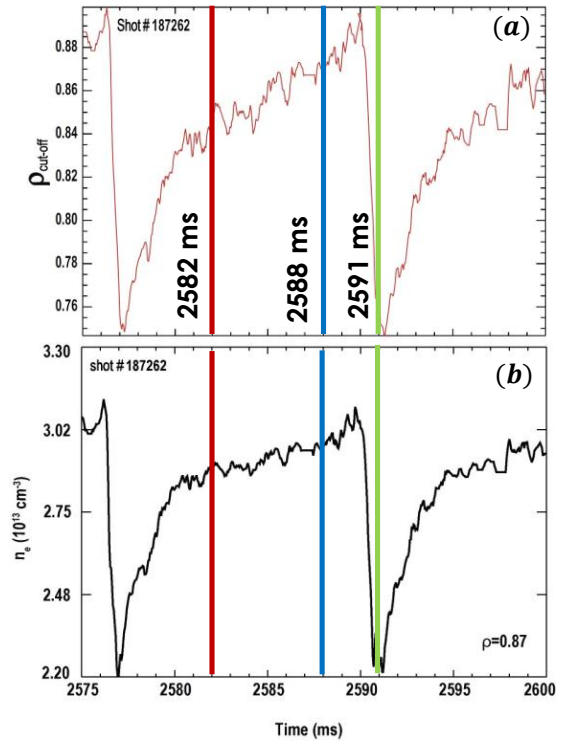

FIG. 11. Reflectometry measurement[29] for 72GHz mm-wave launch in X-mode: (a) DBS cut-off location ($\rho_{cut-off} = 0.87 - 0.75$ for t = 2588 – 2591ms) varies as the ELM changes the cut-off density (b) edge density temporal evolution at radial location ($\rho = 0.87$).

The variability of the broadband fluctuations is potentially caused by changes in coupling of helicon power to the plasma since edge density evolves rapidly during ELM events. Another potentially significant factor is a change in measurement location as the edge density changes. Profile reflectometry measurements[29] are used to determine how edge density evolves and the corresponding changes in the cutoff location for the DBS system. Fig. 11a shows that the cutoff location for a 72GHz DBS beam (X-mode launch) varies from $\rho_{cut-off} = 0.87$ before (2588ms) to $\rho_{cut-off} = 0.75$ after (2591ms) after the ELM crash. Fig. 11b shows the density at $\rho = 0.87$ changes by ~24% between t = 2582 and t = 2591ms.

Notably, Fig. 10c also shows that the amplitude of the sharp peak at 476MHz changes substantially during the period around the ELM, as well. The pattern of evolution is more complex than for the broadband fluctuations with the 476MHz peak varying in a random fashion. A potential explanation for this variability is that the sharp peak is not fully due to pickup, but rather partly due to a plasma wave measurement. This would then lead to interference playing a significant role in the ultimate observed signal level.

## IV. DISCUSSION AND SUMMARY

The principal motivation for development of the new DBS system is to measure the amplitude and spatial distribution of helicon wave power in the plasma (or any slow wave power unintentionally coupled into the plasma edge) during high power helicon injection. These measurements will be used to validate the GENRAY[25] and AORSA[19,28] full-wave models, which predict helicon (and slow wave) propagation, absorption and current drive. This full wave 3D modeling (AORSA) also predicts a complex 3D spatial structure for the helicon wave in the plasma, which makes DBS measurement of the helicon wave challenging. The prototype DBS system described here, and the tests performed with it, establish the feasibility of the concept. The system measures low amplitude broadband fluctuations at frequencies close to the launched helicon wave, at an amplitude ~70dB lower than simultaneously measured turbulence amplitude, although the injected helicon wave is launched toroidally away from the DBS probed location. In particular, the plasma test data demonstrates the sensitivity of the prototype to the helicon fluctuations inside plasma.

The initial plasma test data also offers a valuable opportunity for a preliminary validation effort. For the helicon fluctuation measurements in Fig. 8b and d, the plasma is positioned close to the antenna to ensure strong coupling of the wave to the plasma, leading to strong toroidal localization of the launched wave. Figure 12 shows RF probe measurements of power in the helicon antenna vs. position in shot# 187301 at t= 2001ms. Power fed into the antenna from the end at module # 30 mostly couples to the plasma within the first few antenna modules, which is ~ 1 – 2 parallel wavelengths. This strong localization of the helicon wave at the antenna must be taken into account to understand how power from the antenna spreads through the plasma and reaches the location of the DBS probe beam.

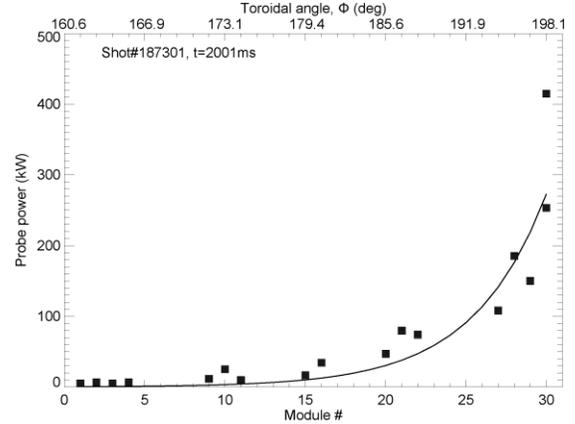

FIG. 12. RF probe measurement of antenna module power at t= 2001ms for plasma shot#187301. Antenna launches power (from $210^0$ port) mostly couples from first few antenna modules at launch end. Solid black line is for projected exponential power fall to estimate antenna power spectrum Vs $N_\parallel$.

The toroidal localization of the launch power distributes the helicon power at the antenna over a broad range of $N_\parallel$ around $N_\parallel \approx 3$. A simple 1D model for the antenna wavefield is constructed to estimate the antenna power spectrum vs. $N_\parallel$ (Fig. 13). Electric field vs. toroidal angle $\phi$ is assumed to be given by $E(\phi) = E_0 \exp(\alpha R_a \phi), \phi \in [0, \Delta\phi_a]$, and $E(\phi) = 0$ otherwise, where $R_a$ is the major radial position of the antenna and $R_a \Delta\phi_a = 1.5$ m is antenna length. The complex value of $\alpha$ is chosen considering the observed drop in wave power in Fig. 12 and a 90° phase-shift between modules. The drop in power is modeled by the solid black line in Fig. 12, which shows power distribution vs position along the antenna assuming a 20% power reduction per module. The modules have a spacing of 0.05 m in the toroidal direction, giving $\alpha = [\ln(0.89) + \frac{i\pi}{2}]/0.05$ m. The spectrum of wave power vs. toroidal index of refraction, $N_\phi = \frac{ck_\phi}{\omega_H}$ (where $\omega_H$ and $c$ are the angular frequency of the launched wave and the speed of light), is given by $\text{pow}_\phi(N_\phi) = \left|\hat{E}\left(\frac{\omega_H}{c}N_\phi\right)\right|^2$, where $\hat{E}(k_\phi)$ is the Fourier transform of $E(\phi)$,

$$\hat{E}(k_\phi) = \left(\frac{1}{2\pi}\right)\oint E(\phi)\exp(-ik_\phi R_a \phi)\,d\phi = E_0\left((\alpha - ik_\phi)R_a 2\pi\right)^{-1}\left(\exp\left((\alpha - ik_\phi)R_a \Delta\phi_a\right) - 1\right).$$

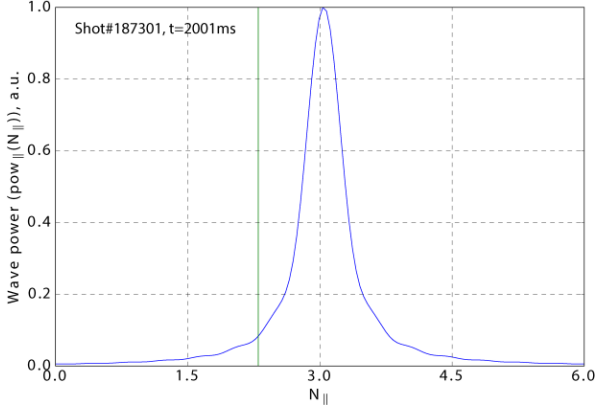

FIG. 13. Modeled power spectrum (shot#187301, t= 2001ms) for the helicon antenna module vs $N_\parallel$. The squared Fourier transform of the projected antenna wave electric field (solid black line from Fig. 14) is shown. Vertical green line marks $N_{//} = 2.3$, corresponding to a helicon ray that would make close approach to DBS ray.

Since, $\text{Re}(\alpha R_a \Delta\phi_a) \approx -3.35$ is a large negative number, $\text{pow}_\phi(N_\phi) \propto \sim \left|\frac{c\alpha}{\omega_H} - iN_\phi\right|^{-2}$. Assuming a field pitch of 15° (the design value for the antenna), the value of $N_\parallel$ for a given value of $N_\phi$ is given by $N_\parallel = N_\phi \cos(15°)$ and the spectrum of power vs. $N_\parallel$ is given by $\text{pow}_\parallel(N_\parallel) = \text{pow}_\phi(N_\parallel/\cos(15°))$. Fig. 13 shows $\text{pow}_\parallel(N_\parallel)$, using the exact equation, normalized to a maximum of 1. The modeled spectrum (Fig. 13) indicates that power is spread over a broad $N_\parallel$ range around $N_\parallel = 3$.

The propagation path for different $N_\parallel$ can be expected to vary significantly. To model which part of the spectrum can reach the DBS probe location, GENRAY ray tracing is used. For helicon waves, the cold plasma dispersion relation is used for the real part of the wavenumber, while a model developed by Chiu, et al.[30] is used for absorption. (GENRAY ray tracing with the Appleton-Hartree dispersion relation is also used to model propagation of the DBS system millimeter wave probe beam as a synthetic diagnostic to interpret the DBS measurements.[2]) A search over multiple values of $N_\parallel$ shows that a ray with $N_\parallel = 2.3$, which has power as high as ~ 8% of the peak power at $N_\parallel \approx 3$ (Fig. 13), can pass very close to the DBS probe location. Fig. 14 shows ray tracing for a helicon ray with index of refraction $N_\parallel = 2.3$ at the antenna. The DBS ray, which propagates at an approximately constant toroidal position of $\phi = 240°$, is also shown in Fig. 14. The helicon ray is marked with a colored circle at every point where it passes $\phi = 240°$. (Note that the helicon ray undergoes toroidal reflections as $N_\parallel$ reverses sign.) Figure 14 shows that on its second pass, the helicon beam passes nearly (separation ~ 5cm) through the DBS beam at cutoff, close enough for a measurement, since the DBS beam has a finite width (2w$_0$ ~10cm)[22] at the cutoff location. Of course, this type of 1D ray tracing model is insufficient to determine the complex spatial distribution of injected helicon beam as well as its interaction location with DBS. Further modeling, for instance with the 3D full wave codes as AORSA[28], is necessary to get a clear idea of helicon-millimeter wave interaction.

In summary, the newly developed DBS system has demonstrated simultaneous measurement of both low frequency turbulence and high frequency fluctuations over a wide RF frequency range. The high frequency channel shows very good ICE sensitivity when using both the non-down-converting (Fig. 2) and down-converting (Fig. 3) receiver circuits. The system has also unambiguously detected high frequency (476MHz) helicon broadband fluctuations during helicon current drive experiments using the down-converting receiver circuit. These broadband high frequency fluctuations are interpreted as resulting from interaction with turbulence near the launch helicon antenna. One likely mechanism for such interaction is modification of the helicon wave coupling through density perturbations modifying the load near the antenna.

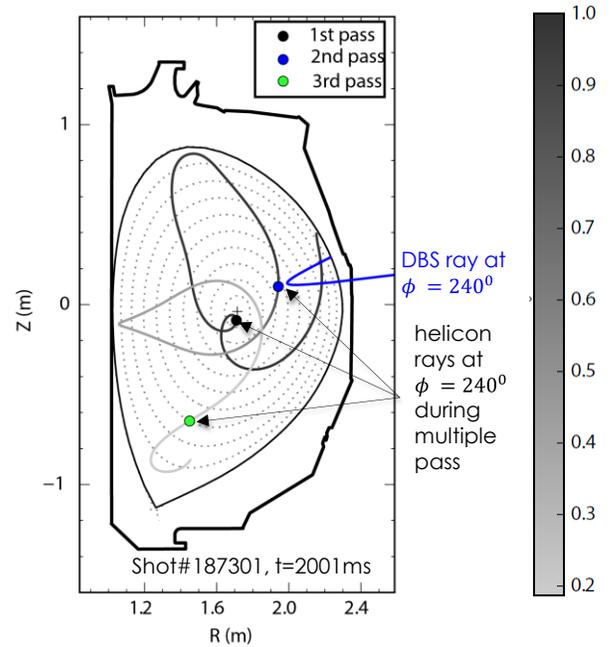

FIG. 14. GENRAY modelling showing helicon and mm-wave propagation throughout the torus. Poloidal view of ray tracing for shot #187301, t = 2001ms: 72GHz, X-mode DBS millimeter wave ray from $\phi = 240°$ (blue) and 476MHz (with $N_\parallel=2.3$) helicon ray launched from $\phi = 180°$ (gray). The helicon ray is marked with a colored circle every time it passes $\phi = 240°$, the location of the prototype DBS. The color of the helicon ray relates to the fraction of unabsorbed power in the ray. At launch, the helicon ray is black, fully unabsorbed, but it becomes lighter as it propagates. For clarity, the ray is terminated when ~ 80% of the initial power is absorbed.

Future upgrades to the system include frequency tunability across the entire E-band range (60-90GHz). This will allow much improved spatial coverage which can even extend past plasma center to the high-field plasma region. In addition, the DBS launch location will be moved a position significantly higher above the midplane than for the plasma tests here. The launch radiation will be quasi-optically coupled into one of the ECH overmoded corrugated waveguides via switch to take advantage of the existing steering capability[31,32]. This geometry will allow improved targeting of the helicon wave. This upgraded E-band DBS system will be able to experimentally investigate helicon wave propagation and, thereby, facilitate determination of wave absorption and the location of current drive. Finally, it will allow detailed comparison with a variety of code predictions for wave propagation (GENRAY[25], AORSA[28]).

## ACKNOWLEDGMENTS

This material is based upon work supported by the U.S. Department of Energy, Office of Science, Office of Fusion Energy Sciences, using the DIII-D National Fusion Facility, a DOE Office of Science user facility, under Award(s) DE-FC02-04ER54698.This work is also supported by U.S. DoE grants DE-SC0020649 and DE-SC0020337. The authors would like to thank Roman Lantsov, Larry Bradley and DIII-D team for their technical support in installing the DBS setup.

## AUTHOR DECLARATIONS

**Conflict of Interest**

The authors have no conflicts of interest to disclose.

## DATA AVAILABILITY

The data that support the findings of this study are available from the corresponding author upon reasonable request.

**Disclaimer:** This report was prepared as an account of work sponsored by an agency of the United States Government. Neither the United States Government nor any agency thereof, nor any of their employees, makes any warranty, express or implied, or assumes any legal liability or responsibility for the accuracy, completeness, or usefulness of any information, apparatus, product, or process disclosed, or represents that its use would not infringe privately owned rights. Reference herein to any specific commercial product, process, or service by trade name, trademark, manufacturer, or otherwise does not necessarily constitute or imply its endorsement, recommendation, or favoring by the United States Government or any agency thereof. The views and opinions of authors expressed herein do not necessarily state or reflect those of the United States Government or any agency thereof.